\documentstyle[12pt,aasms4]{article}
\begin{document}
\title{THE BROAD Fe K LINE PROFILE IN NGC 4151}
\author{Jun-Xian Wang$^{1,2}$, You-Yuan Zhou$^{1,2,3}$, and Ting-Gui
Wang$^{1,2}$}
\altaffiltext{1}{Center for Astrophysics,University of Science and
     Technology of China, Hefei, Anhui, 230026, P. R. China;
     jxw@mail.ustc.edu.cn} 
\altaffiltext{2}{National Astronomical Observatories, Chinese Academy of
     Sciences}
\altaffiltext{3}{Beijing Astrophysics Center, Beijing, 100080, P. R.
     China}  
\begin{abstract}

We present an analysis of the Fe K line profile of NGC 4151 by using long
ASCA observation data obtained in May 1995. 
The unprecedented good data quality, which is much better in the energy
band around 6.4 keV than that of the famous 4.2-day ASCA observation of
MCG --6-30-15 in July 1994, offers a unique opportunity to study the
details of Fe K line profile.
Apart from those characteristics already noticed in earlier ASCA
observations on this object (Yaqoob et al. 1995): a broad and skewed
profile, with a strong peak at about 6.4 keV and a large red wing
extending to $\sim$4 -- 5 keV, which is remarkably similar to that of MCG
--6-30-15, we also find a weak blue wing extending to about 8 keV,
thanks to the good quality of the data. When fitted by a relativistic
accretion disk line plus a narrow core at 6.4 keV, the data constrain the
accretion disk to be nearly face-on, contrary
to the edge-on geometry inferred from optical and UV observations.
However, the extended blue wing can not be well fitted even after we
include corresponding Fe K$\beta$ components. Ni K$\alpha$ line
emission by an amount of 12\% of
Fe K$\alpha$ is statistically required. An alternative explanation is a
model 
consisting of a narrow core and two disk lines with inclinations of
58$^{\rm o}$ and 0$^{\rm o}$, respectively. 
We suppose that the component
with inclination of 58$^{\rm o}$ was observed directly, consistent  
with its edge-on geometry, and the component with inclination of 0$^{\rm
o}$ was scattered into our line of sight by a Compton mirror, which might
be the cool accretion disk corona proposed by Poutanen et al. (1996).
\end{abstract}
 
\section{Introduction}

The fluorescence iron K$\alpha$ line revealed by a 4.2-day ASCA
observation 
in the Seyfert 1 galaxy MCG --6-30-15 has distinct features which are
unique characteristics of the emission line from a relativistic disk
(Tanaka et al. 1995). The line is peaked at around 6.4 keV, with a broad
red wing extending down to 4 keV. The skewness of the profile towards red
and the sharp blue edge near the rest frame energy are a consequence of
the combination of relativistic Doppler effect and strong gravitational
redshift, and are unique to a low-inclination relativistic accretion
disk; other
mechanisms of the line formation are less successful to produce these
features (Fabian et al. 1995). 
However, the alternative models such as thermal Comptonization (Misra and
Kembhavi 1998) can not be ruled out completely. 
Both the disk line fit (Tanaka et al. 1995) and the frequency extrema
analysis (Bromley, Miller \& Pariev 1998) suggested that the inner edge of
line formation region is within a few Schwarzschild radii, providing 
ever strongest evidence for a supper-massive black hole in this object. 
Similar broad line profiles with lower statistics have also been seen in
other AGNs (Mushotzky et al. 1995, Tanaka et al. 1995, Yaqoob et al. 1995,
Nandra et al. 1997, Turner et al. 1998 and references therein). 

The line profile is strongly dependent on the inclination of the
relativistic accretion disk (Fabian et al. 1989), thus providing a method
to estimate the inclination of inner accretion disk from line profile
fitting. The inclinations from this method for a sample of Seyfert 1
galaxies are strongly constrained to be nearly face-on (Nandra et al.
1997), consistent with the expectation of the unification model of the
two types of Seyfert galaxies (see Antonucci 1993 for a review). However,
there are complications to this simple picture. Turner
et al. (1998) found that type 2 Seyfert galaxies also possess similar
iron K$\alpha$ line profiles, indicative of face-on accretion disk as
well.
This might be due to the strong contribution of a narrow component in
these objects (Weaver \& Reynolds 1998) or due to complex absorption (Wang
et al. 1999a). Moreover, the rapid variability of the line equivalent
width (EW)
and profile observed in MCG --6-30-15 (Iwasawa et al. 1996) and NGC 4051
(Wang et al. 1999b) cannot be readily explained by any current simple 
accretion disk line model.

NGC 4151 is a bright nearby (z = 0.0033) Seyfert 1.5 galaxy. The edge-on
orientation of its nucleus is strongly supported by the biconical geometry
of the [O III] $\lambda$5007 region (Evans et al. 1993, Pedlar et al.
1993) and the best estimated inclination is 65$^{\rm o}$.
Yaqoob et al. (1995) presented the first measurement of its broad Fe K
line profile from the ASCA observations performed in 1993
May, November and December. The apparent line profile is asymmetric,
consisting of a peak at $\sim$6.3 -- 6.4 keV and a broad red wing
extending to $\sim$4 -- 5 keV. When a disk line model is fitted, the
inclination angle of the disk ($\theta$ = 0$^{+19}_{-0}$ degrees) is
strongly constrained to be face-on, in contrary to the edge-on geometry of
this source. This problem can be eased, but not solved, by assuming an
additional narrow-line contribution, presumably from the torus, to the Fe
K line ($\theta$ = 25$^{+3}_{-8}$ degrees, see Yaqoob \& Weaver 1997).

A much longer ASCA observation (3 days) of NGC 4151 was carried out in
1995, which provides data with better statistics (Leightly et al. 1997). 
In this letter, we report measurement of the complex broad iron K
line profile and its implication.

\section{The ASCA Data} 
NGC4151 was observed by ASCA from 1995 May 10 to 12 with the Solid-state
Imaging Spectrometer (SIS) in 1CCD mode and the Gas Imaging Spectrometer
(GIS) in PH mode. 
The GIS data are contaminated by a nearby BL Lac object due to their worse
spatial resolution.
In this letter, we will concentrate on the SIS data, also because
they offer much better energy resolution (Inoue 1993), which is crucial to
the Fe K line profile analysis.
The data were reduced with the ASCA standard software
XSELECT and cleaned using the following screening criteria: satellite not
passing through the South Atlantic Anomaly (SAA), geomagnetic cutoff
rigidity greater than 6 GeVc$^{-1}$, and minimum elevation angle above
Earth's limb of 10$^{\rm o}$ and 20$^{\rm o}$ for nighttime and daytime
observations, respectively. 
Source counts were extracted from a circular area of radius 3.4'
for the SIS0 and SIS1, and the background counts were
estimated from the
blank sky data. Spectra extracted from SIS0 and SIS1 were combined
together and grouped to have $>$20 counts in each bin to
allow the use of $\chi^2$ statistics. Spectral analysis was carried out
using XSPEC.

The ASCA observation lasted for about three days, and the SIS0 detector
received a net exposure time of 93,000s and an average count rate of 1.49
cts/s in the 0.4 -- 10.0 keV band, while these two parameters for the
well-known 4.2-day ASCA observation on MCG --6-30-15 in 1994 July are
150,000s and 1.82 cts/s, respectively. Although shorter exposure time and
lower count rate may probably lead to worse statistics, we noticed that,
because of its much harder spectrum (due to strong absorption), the actual
total counts (32,000) for NGC 4151 in the 5.0 -- 7.0 keV band are almost
three times of those (11,000) for MCG --6-30-15 in the same energy band. 
Thus,
the NGC 4151 Fe K spectrum used in this paper has better statistics than
the average MCG --6-30-15 spectrum and a best ever quality Fe K line
profile. 

\section{Spectral fits}
Following Weaver et al. (1994), we fit the underlying continuum in the 
1.0 -- 4.0 and 8.0 -- 10.0 keV bands (to exclude the possible broad iron 
line region) with a model which consists of a dual
absorbed power law with some fraction (a best-fit value for this letter
is $\sim$5\%) of the direct continuum scattered into our line of sight and  
absorbed only by the Galactic column of 2 $\times$ 10$^{20}$ cm$^{-2}$. 
We do not include a Compton reflection component in the spectral fits
because a clear reflection component has never been detected
(Maisack \& Yaqoob 1991; Yaqoob et al. 1993). Zdziarski et al. (1996)
claimed that they detected a reflection component. However, 
considering the poor statistical significance of
their detection (they used only a delta-chi square of 2.7 for the their
errors yet the model was complicated), the complex intrinsic continuum
which was unknown, and the cross-calibration of two different satellites
(Ginga and GRO/OSSE) which is not perfect, we think that their result is
questionable.   

Our model can describe the data reasonably well ($\chi^2$ = 364 for 325
degrees of freedom), and the best-fit model parameters converge to
$\Gamma$ = 1.30$\pm{0.07}$, N$_H$(low) $\simeq$ 3.2 $\times$ 10$^{22}$
cm$^{-2}$ (covering $\sim$ 40\% of the source), and N$_H$(high) = $\sim$
10.9 $\times$ 10$^{22}$ cm$^{-2}$ (covering $\sim$ 60\% of the source). 
The index is slightly flatter than that seen previously for NGC 4151 as
derived from data weighted toward higher energies (Yaqoob et al. 1993). We
have also tried a model consisting of a power-law absorbed by ionized
material (Zdziarski et al. 1995) plus a fraction of scattered underlying
continuum, but failed to get a satisfactory fit ( $\Gamma$ = 1.00,
$\chi^2$ = 393 for 326 degrees of freedom).

The profile of the iron K line is shown
in Figure 1a, which is similar to, and much better defined than, the
one presented by Yaqoob et al. (1995). The line shows a strong narrow
peak around 6.4 keV and a huge red wing, containing a second slightly weak
peak at around 5.2 keV, extending to $\sim$ 4.5 keV. In this respect, the
line profile of NGC 4151 is remarkably similar to that seen in MCG
--6-30-15 (see Figure 1b for comparison). We also find a weak blue
wing extending to about 8 keV, which is just visible in Figure 2c of
Yaqoob et al. (1995), is now clearly seen, thanks to the greatly
improved statistics. The brightness of this source rules out the
possibility that the blue wing is caused by improper background
subtraction (only 1$\%$ of the count rate at 7. -- 8. keV is due to the
background).

First, we fit the line with a disk-line model (Fabian et al. 1989) plus a
narrow core at 6.4 keV which is presumably from the torus and also
statistically required ($\Delta\chi^2$ = -46). 
Though expected in theory, the fluorescence Fe K$\beta$ line is
always ignored when fitted to data with much lower quality.
Considering the high quality of the data in this letter, we also include
the corresponding Fe K$\beta$ components (disk-line plus narrow core)
in all of our models by amounts of 11.3\% of Fe K$\alpha$ (George \&
Fabian 1991). In fact, Fe K$\beta$ components are also statistically required
here ($\Delta\chi^2$ = - 36).
The outer radius ($R_o$) of
the disk is fixed at 1000 $R_g$ ($R_g$ = GM/c$^2$) to minimize the numbers
of free parameters because the fits are not sensitive to the value
for the outer radius.
The best fit value is also consistent with 1000 $R_g$ when it is allowed
to vary. Considering the strong peak at 6.4 keV, we fixed the disk-line
energy at 6.4 keV in the source rest frame. Results of this fit are given
in Table 1 (Model A). The inclination of the disk is 27$^{\rm o}$, similar
to the value obtained by Yaqoob \& Weaver (1997).
However, the blue-wing of the line beyond the 6.8 keV has not been well
fitted (see Figure 2a). 

In the accretion disk model context, the high energy blue wing can be
produced by an accretion disk with high inclination angle. The blue wing,
as well as the edge-on geometry of NGC 4151 inferred from optical and UV
observations (Evans et al. 1993), motivated us to add an extra component
of a disk line with large inclination to Model A (Model B, Fe K$\beta$
components also included). We assume the
inner and outer radii of the disk for these two disk-line components are
the same, while the emissivity index {\rm q} (F$_{line}\propto R^{-q}$) is
allowed to vary independently considering the fact the line photons from
different parts of the disk may have different scattering fraction.
The results are also given in Table 1. The model can fit the observed line
profile fairly well (see Figure 2b). The improvement of Model B to Model A
is significant ($\Delta\chi^2$ = -19). As expected, the second disk
component requires a high inclination of $58_{-12}^{+32}$ degrees, while
the first component requires a face-on disk ($0^{+12}_{-0}$ degrees).

Alternatively, the high energy blue wing between 6.8 and 8.0 keV may also
be produced by the Ni K$\alpha$ line emission at 7.48 keV. In order to 
reduce the excesses in the residuals between 6.8 and 8.0 keV in Figure 2a,
corresponding Ni K$\alpha$ components by amounts of 12$^{+6}_{-6}$\% of
Fe K$\alpha$
have to be added to Model A (see Figure 2c, $\Delta\chi^2$ = -13). 

\section{Discussion}
Zdziarski et al. (1996)
argued that the ASCA data of NGC 4151 obtained in May 1993 can be modeled
by complex absorption and a narrow Fe K line only, with no broad line
required, contrary to the results of Yaqoob et al. (1995) and this letter. 
However, the ASCA data used by Zdziarski et al have much lower
signal-to-noise.
Through fitting the unprecedented high quality spectra of NGC 4151
obtained by ASCA in 1995, we find a prominent broad Fe K line profile in
NGC
4151 (see Figure 1), which is remarkably similar to that of MCG
--6-30-15.  
A model consisting a dual absorbed power law plus a scattered component
and a narrow Fe K line at 6.4 keV ($\sigma$ = 0.11$^{+0.02}_{-0.02}$) fits
the data (1.0 -- 10.0 keV) poorly ($\chi^2$ = 706 for 598 degrees of
freedom) and results broad systematic positive residuals around 6.0 keV
(see Figure 2d).
An extra broad line is statistically required ($\Delta\chi^2$
= -78).

The high signal-to-noise profile of NGC 4151 obtained in this letter
shows a strong narrow peak around 6.4 keV and a huge red wing, containing
a second slightly weak peak at around 5.2 keV, extending to $\sim$ 4.5
keV. In this respect, the line profile of NGC 4151 is remarkably similar
to that seen in MCG --6-30-15.
We also find a weak blue wing extending to about 8.0 keV, which is
just visible in Figure 2c of Yaqoob et al. (1995), is now clearly
seen (Figure 1a), thanks to the greatly improved statistics. There may
also be such
weak blue wing in the Fe K profile of MCG --6-30-15 (see Figure 1b). 

When fitted by the disk-line model plus a narrow core, a face-on disk is
also required ($\theta$ = 27$^{\rm o}$), contrary to the edge-on geometry 
inferred from optical and UV observations (Evans et al. 1993, Pedlar et
al. 1993). We consider that the Compton mirror of Poutanen et al. (1996),
who proposed that the central source of NGC 4151 is completely hidden from
our line of sight and observed X-ray radiation is produced by
scattering in the higher, cooler parts of the accretion disk corona, or
in a wind, could solve the problem well.
We also want to point out that scattering by
an "ionizing cone" would also explain the similarity of Fe K line profiles
seen in Seyfert 2 galaxies and in Seyfert 1 galaxies (Turner et al. 1998),
indicating reprocessing by face-on disks and contradicting the expectation
of the unification scheme. The variability properties
of the Fe K line are important to test such scattering hypothesis.

Besides the broad profile of the Fe K line, the blue wing extending to
about 8.0 keV (see Figure 1a), which could not be well fitted by a simple
disk line plus a
narrow core even after the corresponding Fe k$\beta$ components
included, should also be paid much attention to. One possible
explanation is that we observe not only a broad Fe K line in
NGC 4151, but also a corresponding Ni K$\alpha$ line by an amount of
12$^{+6}_{-6}$\% of Fe K$\alpha$. 
This may be the first detection of Ni K$\alpha$ fluorescence line
emission in the X-ray spectrum of AGNs.

However, we notice that the energy of Ni K$\alpha$ line is just above that
of the Fe K absorption edge. The line flux, which is sensitive to the
Ni abundance and X-ray spectral slope, can be substantially reduced due to
photo-electron absorption by Fe atoms. So, the contribution of Ni
K$\alpha$ emission line might be much weaker than expected. 
An alternative explanation to the weak blue wing is an extra Fe K disk
line with high inclination angle (it is also possible that both the Ni
K$\alpha$ components and the extra disk line are important in reality).
Poutanen et al. (1996) assumed that the central source is completely
hidden from our line of sight by the thick part of the accretion disk wind
or by the broad emission-line clouds. However, it is plausible to suppose
that the thick material covers the X-ray only partially, and part of the
central continuum is seen directly. The fast variability and the
complex absorption of the X-ray continuum seem to support this idea.
The results of two-disk component model fit (Model B) is fully consistent
with this
picture. Poutanen et al. (1996) assumed that the scattering region is
situated along the axis of the accretion disk in the form of a cone. The
half opening angle  of the "ionized cone" determined from the optical and
UV observation is about 35 degree (see Fig. 10 of Evans et al. 1993), the
angle-averaged line profile in the scattered light should resemble the one
from a face-on disk, which corresponds to the low inclination line
component in our two disk-line fit. The direct component should show the
same inclination as the system, which is 65$^{\rm o}$ inferred from
optical and UV observations. We got an inclination of 58$_{-12}^{+32}$
degrees for this line component, which is in good agreement with
interpreting it as a direct component. Comparable EW of the two line
components suggests that the flux of the X-ray scattered should be
commensurate roughly with the directly observed flux.
The little lower parameter $q$ of the direct disk-line indicates that
more direct Fe K photons can be observed from large disk radii.

\acknowledgments
This work is supported by Chinese National Natural Science Foundation,
PanDeng Project and Foundation of Ministry of Education. We thank the
referee for many critical comments, especially on the possility of Ni
K$\alpha$ contribution, which significantly improve the presentation of
this paper.

\newpage 
\input psfig.sty
\begin{figure}
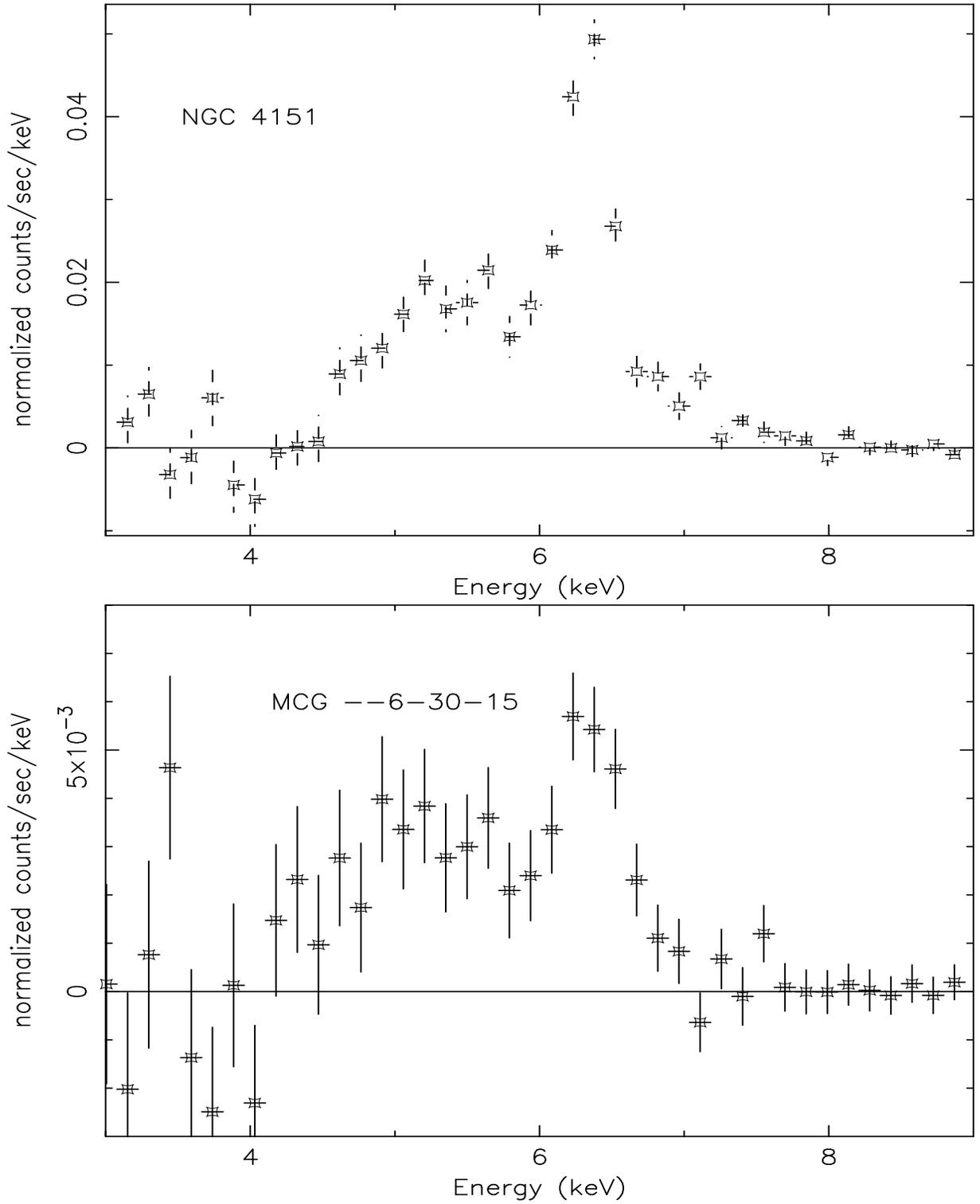

\centering
\psfig{figure=f1a.ps,width=16.2cm,height=10.0cm,angle=-90}
\psfig{figure=f1b.ps,width=16.2cm,height=10.0cm,angle=-90}
\figcaption[f1a.ps, f1b.ps]{(a) The high S/N ratio Fe K line profile in
NGC 4151. (b) The profile of Fe K line in MCG --6-30-15 (from SIS data
of the well-known 4.2 days ASCA observation in 1994). It's
obvious that the profile of Fe K line in NGC 4151 has a much 
higher S/N ratio.\label{fig-1}}
\end{figure}

\clearpage
\newpage
\begin{figure}
\psfig{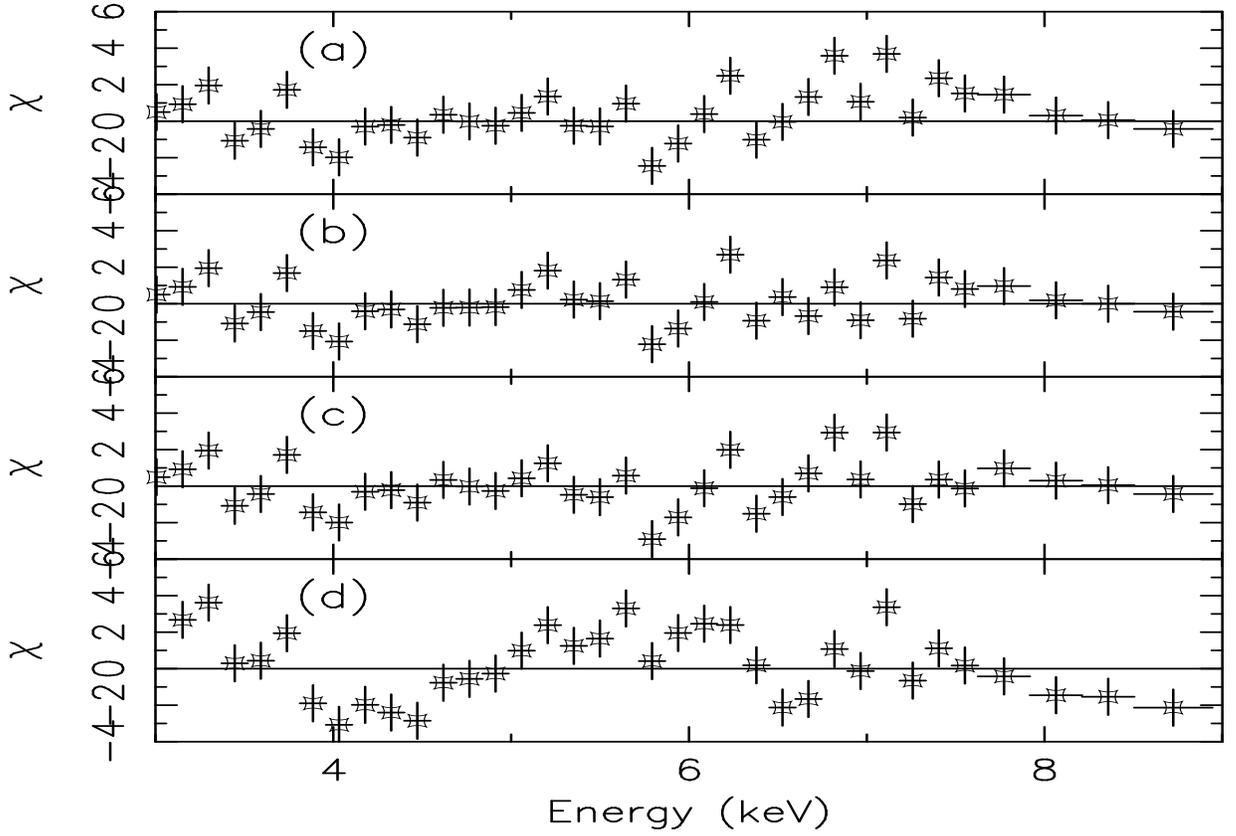}
\figcaption[f2.ps]{Plots of residuals for models:
(a) a single cold disk-line plus a narrow core (Model A); (b)
two cold Fe K disk-line with different inclinations plus a narrow core
(Model B); (c) including corresponding Ni K$\alpha$ components by amounts
of 12\% of Fe K$\alpha$ to Model A. Fe K$\beta$ components by amounts of
11.3\% of Fe K$\alpha$ are included in both Model A and Model B; (d)
a model consisting a dual aborbed power law plus a scattered component and
a narrow Fe K line at 6.4 keV. \label{fig-2}}
\end{figure}

\newpage
\begin{table}  
\caption[]{Fe K line fits$^a$}  
\begin{tabular}{cccccccc}
\\
\hline \hline
Model  & $\chi^2/dof$ & $R_i$ ($R_g$) & EW$_{core}$ (eV) & EW$_{disk}$
(eV) & $q$ & $\theta$ 
\\ 
\hline  
A & 628/599 & 8.9$^{+0.9}_{-0.9}$ & 71$^{+13}_{-12}$ & 276$^{+21}_{-24}$ &
-4.2$^{+0.7}_{-0.8}$ & 27$^{+1}_{-2}$\\

B & 609/596 & 6.0$^{+0.8}_{-0.0}$ & 89$^{+11}_{-10}$ & 136$^{+28}_{-27}$ &
-2.0$^{+0.4}_{-0.3}$ & 58$^{+32}_{-12}$\\
  &         &                     &                  & 150$^{+18}_{-19}$
&
-3.3$^{+0.3}_{-0.2}$ & 0$^{+12}_{-0}$\\

\hline
\end{tabular}\\
\\
$^a$Errors are given at 90\% level. The energy of narrow core is fixed at
6.4 keV in the galaxy rest frame and the line width fixed at zero.
Model A: a single cold disk line (6.4 keV in the galaxy rest frame) plus a
narrow core; Model B: two cold disk line with different inclinations
plus a narrow core. Fe K$\beta$ components by amounts of 11.3\% of Fe 
K$\alpha$ are included in both models.
\end{table}

\end{document}